\documentclass[usenatbib]{mnras}

\usepackage[utf8]{inputenc}
\usepackage[T1]{fontenc}
\usepackage{ae,aecompl}

\usepackage{graphicx}
\usepackage{float}
\usepackage{amsmath}
\usepackage{amssymb}
\usepackage{enumitem}

\graphicspath{{figs/}}  

\def\beq{\begin{eqnarray}}
\def\eeq{\end{eqnarray}}

\def\zhalf{z_{1/2}}
\def\zlmm{z_{lmm}}
\def\cvir{c_{vir}}

\def\mpeak{M_{peak}}
\def\vmax{V_{max}}
\def\vpeak{V_{peak}}
\def\zpeak{z_{peak}}
\def\mhalo{M_{h}}

\def\hmsun{h^{-1} \; \mathrm{M_{\odot}}}
\def\h{h^{-1}}
\def\Dn{D_{n}}  
\def\mstar{M_{\star}}

\def\Nrun{48}

\title[Secondary and Neighbour Bias]{Spatial Clustering of Dark Matter Haloes: Secondary Bias, Neighbour Bias, and the Influence of Massive Neighbours on Halo Properties}

\author[A. N. Salcedo et al.]{Andr\'{e}s N. Salcedo$^{1}$\thanks{E-mail: salcedo.11@osu.edu}, 
Ariyeh H. Maller$^{2,3}$\thanks{E-mail: amaller@citytech.cuny.edu},
Andreas A. Berlind$^{4}$,
Manodeep Sinha$^{5,6,4}$, 
\newauthor Cameron K. McBride$^{4,7}$,
Peter S. Behroozi$^{8}$, 
Risa H. Wechsler$^{9}$, 
and David H. Weinberg$^{1}$
\\
$^{1}$Department of Astronomy and Center for Cosmology and AstroParticle Physics, The Ohio State University, Columbus, OH 43210, USA \\
$^{2}$Department of Physics, New York City College of Technology, CUNY, 300 Jay St., Brooklyn, NY 11201, USA \\
$^{3}$Department of Astrophysics, American Museum of Natural History, New York, NY, USA \\
$^{4}$Department of Physics and Astronomy, Vanderbilt University, 1807 Station B, Nashville, TN  37235, USA \\
$^{5}$Centre for Astrophysics and Supercomputing, Swinburne University of Technology, Hawthorn, Victoria 3122, Australia \\
$^{6}$ARC Centre of Excellence for All Sky Astrophysics in 3 Dimensions (ASTRO 3D) \\
$^{7}$Harvard-Smithsonian Center for Astrophysics, 60 Garden St., Cambridge, MA 02138, USA \\
$^{8}$Department of Astronomy and Steward Observatory, University of Arizona, 933 N. Cherry Ave., Tucson, AZ 85719 USA\\
$^{9}$Physics Department, Stanford University, Department of Particle and Particle Astrophysics, SLAC National Accelerator Laboratory, \\ Kavli Institute for Particle Astrophysics and Cosmology, Stanford, CA 94305, USA}

\date{Accepted 2018 January 05. Received 2017 December 22; in original form 2017 September 11.}

\pubyear{2017}

\begin{document}
\label{firstpage}
\pagerange{\pageref{firstpage}--\pageref{lastpage}}
\maketitle

\begin{abstract}
We explore the phenomenon commonly known as halo assembly bias, whereby dark matter haloes of the same mass are found to be more or less clustered when a second halo property is considered, for haloes in the mass range $3.7 \times 10^{11} \; \hmsun - 5.0 \times 10^{13} \; \hmsun$. Using the Large Suite of Dark Matter Simulations (LasDamas) we consider nine commonly used halo properties and find that a clustering bias exists if haloes are binned by mass or by any other halo property.  This {\emph{secondary bias}} implies that no single halo property encompasses all the spatial clustering information of the halo population. The mean values of some halo properties depend on their halo's distance to a more massive neighbour.  Halo samples selected by having high values of one of these properties therefore inherit a \emph{neighbour bias} such that they are much more likely to be close to a much more massive neighbour.  This neighbour bias largely accounts for the secondary bias seen in haloes binned by mass and split by concentration or age. However, haloes binned by other mass-like properties still show a secondary bias even when the neighbour bias is removed. The secondary bias of haloes selected by their spin behaves differently than that for other halo properties, suggesting that the origin of the {\emph{spin bias}} is different than of other \emph{secondary biases}. 
\end{abstract}



\section{Introduction}

Halo assembly bias is the phenomenon found in cosmological N-body simulations that the clustering of dark matter haloes depends on halo properties other than mass. Detected by \citet{st:04} and \citet{gsw:05} for a measure of halo age it was soon realized that this clustering dependence not only exists for various measurements of halo age \citep{gsw:05,wech:06,gw:07,wmj:07, lmg:08} but also for concentration \citep{wech:06,gw:07,fw:10,lazeyras:17,villarreal:17}, spin \citep{gw:07,fw:10,lp:12,lazeyras:17,villarreal:17}, halo shape \citep{fw:10,lazeyras:17,villarreal:17} and the amount of substructure in the halo \citep{wech:06, gw:07}. These varying measurements have collectively been termed assembly bias \citep{croton:07} because the original result involving age showed that halo clustering is biased with respect to halo assembly history. Thus it is often assumed that the reason other halo properties show clustering dependences is because those properties are correlated with the halo's assembly history. However, some properties that do not correlate strongly with assembly history (e.g., spin) display a strong bias signal at fixed mass, while other properties that are directly related to the history (e.g., the scale of the last major merger) do not \citep{lmg:08}, making it unclear if that assumption is warranted. 

The reason halo assembly bias is of interest, besides for understanding the growth and properties of dark matter haloes, is that it questions assumptions that have traditionally been made by statistical models that connect the clustering of dark matter to the clustering of galaxies. There are a number of methodologies used to make this connection, such as the halo occupation distribution \citep[HOD; e.g.,][]{ps:00,sshj:01,bw:02,cs:02,berl:03,zm:15,zm:16}, the conditional luminosity function \citep[CLF; e.g.,][]{ymv:03,vym:03} and subhalo abundance matching \citep[SHAM; e.g.,][]{vo:04,cwk:06}. Regardless of the methodology used, all these techniques share the common feature that they connect galaxies to dark matter haloes through a simple parameterization instead of a full physical model of galaxy formation. The parameterization is then tested by comparing galaxy clustering in the model to the observed clustering of galaxies (and/or galaxy-galaxy lensing).  Traditionally, the parameterization of these models is based solely on halo mass (or a single mass-like parameter) in part because it was believed that the clustering of dark matter haloes only depends on their mass (although SHAM models also implicitly account for the bias with respect to substructure). Halo assembly bias undermines the rationale for this assumption, questioning the suitability of the entire approach.

Galaxies are connected to dark matter haloes in some way, thus there is some HOD/CLF/SHAM model that correctly describes our universe. The main issue then is how complicated is that connection. Most SHAM models allow for mass stripping of subhaloes as they move through a halo by a simple global offset for subhalo masses \citep{vo:06,wcdk:08}, by using halo masses at the time of accretion \citep[e.g.][]{cwk:06,swdfko:12}, or by using another quantity such as peak circular velocity that is less strongly affected by tidal stripping \citep[e.g.][]{rwtb:13}. The following three classes of galaxy-halo models can be used to describe their different levels of complexity. 


\begin{itemize}[leftmargin=*]
\item \emph{traditional} - the statistics of a halo's galaxy population depends only on halo mass. This applies to most HOD and CLF models, as well as halo abundance matching models that ignore subhaloes.
\item \emph{simple} - a halo's galaxy population depends on only one halo property, but it is not necessarily mass. This applies to most SHAM models.
\item \emph{complex} - a halo's galaxy population depends on more than one halo property. This applies to extensions to the HOD or SHAM, like the `decorated HOD' \citep{hearin:16}, `environment dependent HOD' \citep{mcewen:16}, `age matching' models \citep{hw:13} or the generalized SHAM model with adjustable concentration dependence of \citet{lehmann:17}.
\end{itemize}

With this language we see that one issue raised by halo assembly bias is whether the galaxy-halo connection can be parameterized in a simple manner or whether a more complex parameterization is required. We emphasize that the existence of halo assembly bias does not in and of itself contradict the assumption of traditional HOD, CLF or SHAM.  It is possible that halo clustering depends in a complex way on many halo properties, but that the galaxy population in a halo only depends on halo mass.  Analytic or semi-analytic models of galaxy formation suggest that other halo properties besides mass play a secondary role in the galaxy formation process, but it is possible that more complex treatments find that this is not the case.  Alternatively, it is possible that some halo properties affect some galaxy properties but leave others unchanged.  For example, halo formation time may affect a galaxy's current star formation rate, but not its total mass in an appreciable way.  Moreover, halo properties may affect galaxy properties but to an extent that is too small to change the clustering of galaxy properties at a measurable level.  Whether an HOD or other methodology needs to be traditional, simple or complex is also a function of the data set and the questions being asked.  

In this paper we explore halo assembly bias in a large suite of cosmological simulations with a goal of better understanding its causes.  We generalize the concept of assembly bias by not just considering how the clustering of haloes in a bin of mass depends on other halo properties, but how the clustering of haloes binned by any halo property depends on other halo properties. We call this `secondary bias' \citep[][]{mzw:17}. We then study how halo properties depend on a halo's distance from a more massive neighbour. This dependence implies that when haloes are selected by some secondary properties, a `neighbour bias' is created whereby these haloes are much more likely to be near more massive haloes.  We show that if this neighbour bias is controlled for, then haloes binned by mass have almost no secondary bias with age or concentration. 

The paper is organized as follows. In Section \ref{sec:sims} we describe our simulations and halo catalogs. In Section \ref{sec:meth} we discuss our methodology for studying secondary bias. In Section \ref{sec:split} we look at generalizations of assembly bias binning and then splitting haloes based on different halo properties.  In Section \ref{sec:Mc} we explore how halo properties depend on distance from a massive neighbour and how this dependence is related to clustering biases.  We conclude in Section \ref{sec:conc}.


\section{The Simulations}
\label{sec:sims}

We study simulated haloes from the Large Suite of Dark Matter Simulations (LasDamas) project\footnote{http://lss.phy.vanderbilt.edu/lasdamas/}. LasDamas consists of a series of cosmological N-body simulations run in four boxes of varying sizes and mass resolutions. All boxes in the current phase of LasDamas use a $\Lambda$CDM cosmological model based on the {\sc{planck}} satellite's measurements \citep{planck:14}: $\Omega_{m}=0.302, \Omega_{\Lambda}=0.698, \Omega_{b}=0.048, h=0.681, \sigma_{8}=0.828, n_{s}=0.96$.

In this paper we focus on one of the highest resolution boxes, \emph{Consuelo}. This run is a periodic cube with a side length of $L=420h^{-1}$Mpc that contains $N_{p}=1400^{3}$ particles of mass, $m_{p}=1.87\times10^{9} \; \hmsun$.  The gravitational force softening is $\epsilon_{g}=8h^{-1}$kpc.  We use $\Nrun$ realizations of this box with different initial perturbations, which gives us a much larger effective volume and allows us to measure box to box variations.

The initial power spectrum of density fluctuations was computed using \emph{CAMB} \citep{aa:11}. An initial density field at $z=99$ was generated and initial positions and velocities computed for the particles using the \emph{2LPT} code \citep{scoc:97}. \emph{2LPT} computes initial conditions using second-order Lagrangian perturbation theory. This method is more accurate than the traditional Zel'dovich approximation because it accounts for very early non-linear gravitational evolution, which can have a significant impact on the properties of the highest density peaks. \emph{2LPT} initial conditions have been tested extensively by \citet{scps:06}.
Once the initial positions and velocities were generated, the gravitational evolution was performed using the publicly available \emph{Gadget-2} code \citep{spri:05}. We only used collisionless dark matter particles and utilized its TreePM functionality to speed computation and increase the long range force accuracy.

We used the software package {\sc{rockstar}} version 0.99.9 \citep{bww:13} to identify haloes. The haloes were defined as spherical over-densities with a mean virial density as defined by \citet{bn:98} and unbound particles were not removed. Merger trees were created using the {\sc{consistent trees}} package \citep{behr:13} tracking each halo's history through the simulation. 


In this paper we make use of the following halo properties at redshift zero.
\begin{itemize}[leftmargin=*]
\item
The halo mass $\mhalo$.
\item 
The maximum circular velocity $\vmax$. 
\item
The halo concentration $\cvir$, measured using the method described by \citep{klypin:11}. This uses measurements of the halo's maximum circular velocity and virial radius instead of fitting a NFW profile \citep{nfw:96} to the halo density distribution because those quantities can be measured more robustly.  Then through linear interpolation one solves:
\begin{equation}
\frac{c_{vir}}{f(c_{vir})} = \vmax^2 \frac{R_{vir}}{G \mhalo} \frac{2.1626}{f(2.1626)}
\end{equation}
where $f(x)$ is given by:
\begin{equation}
f(x) \equiv \ln (1 + x) - \frac{x}{1 + x}.
\end{equation}
If the halo density profile is well fit by an NFW then this will give identical results as a fit to the profile, but for haloes poorly fit by the profile this method gives more reasonable values of $\cvir$. We have found in the course of this work that an early version of {\sc{rockstar}} gave a different value for the relative bias of haloes when split by NFW-fitted concentrations. However the bias when split by $\vmax$ based concentration is consistent across versions of {\sc{rockstar}}. Thus one is cautioned that the details of how haloes are found and properties are fit may play a role in this type of analysis.
\item
The halo spin, $\lambda$, is calculated as defined in \citet{bull:01}:
\begin{equation}
\lambda=\frac{\left\Vert J\right\Vert }{\sqrt{2} \mhalo V_{vir}R_{h}}
\end{equation}
where $J$ is the halo angular momentum and $V_{vir}$ is the circular velocity at the halo's virial radius, $R_h$. 
\end{itemize}
Furthermore, since we have merger trees that trace each halo back through all previous saved time steps, we also have the following properties of halo histories.
\begin{itemize}[leftmargin=*]
\item
The peak mass  $\mpeak$, which is the highest value of $\mhalo$  in a halo's history.
\item
The peak maximum circular velocity $\vpeak$, which is the highest value of $\vmax$ in a halo's history.
\item
The redshift at which the halo first achieves $\mpeak$, which we denotes as $\zpeak$.  Note that if a halo stops growing, then $\zpeak$ will be the redshift where it first attains its $z=0$ halo mass. 
\item
The redshift when a halo had its last major merger (a merger of ratio 1:3 or greater), which we denote by $\zlmm$.
\item
The halo's dynamical time averaged accretion rate, $\dot{M}_{\tau_{dyn}}$. A halo's dynamical time is defined by:
\begin{equation}
\tau_{dyn} = \frac{1}{\sqrt{G \rho_h}} \sim 3 \; \mathrm{Gyr}.
\end{equation}  
Because of the spherical over-density halo definition this is a mass-independent property.
\item
The halo age, defined as the redshift at which the halo accreted half of its {\emph{peak mass}}, which we denote by $\zhalf$. 
\end{itemize}

We only study distinct haloes; no subhaloes (defined as haloes whose centres are within another halo's $R_{vir}$) are included in any of the analysis.  

\begin{table}
\caption{The total number of haloes in our 48 simulation boxes that are above the cut used for different samples. We consider samples determined by the $z=0$ halo mass and maximum circular velocity, as well as by the peak halo mass or circular velocity in a halo's history.}
\label{tab:cut}
\begin{tabular}{lcc}
\hline
Halo Property  & Cut Used & Number of haloes \\
\hline
      $\mhalo$        & $\mhalo > 3.74 \times10^{11}\; \hmsun$ & 33.1 million\\
      $\vmax$         & $\vmax > 130 \; \mathrm{km \; s^{-1}}$  & 31.7 million \\
      $\mpeak$       & $\mpeak > 5.0 \times10^{11} \; \hmsun$ & 30.0 million\\
      $\vpeak$        & $\vpeak  > 160 \; \mathrm{km \; s^{-1}}$ & 25.4 million \\
\hline
\end{tabular}
\end{table}

While some halo properties, like mass, are well determined with just a hundred particles, other properties may require many more particles before they can be reliably determined.  \citet{klyp:15} claim that the mass and velocity functions converge with only 50 particles per halo. \citet{onor:14} show that for many properties one needs 1000 particles to ensure that the properties only exhibit small changes when they are resimulated at higher resolution, while for halo spin this is as much as 10,000 particles \citep[][]{benson:17}. However, for this study it is not critical that we accurately resolve these quantities because we are for the most part just using them to make high and low quartile subsamples. For example, suppose we wish to identify the highest quartile of concentrations in a mass bin. While errors in measuring the concentration may scatter some haloes in and out of this quartile, as long as the errors are smaller than the intrinsic spread in concentration this will not affect a large fraction of haloes.  Errors will only cause a  weakening of the assembly bias signal.  We consider haloes with at least 200 particles when using a halo mass or $\vmax$ selected sample.  This corresponds to a minimum halo mass of $\mhalo \ge 3.74 \times 10^{11} \; \hmsun$ or a minimum maximum circular velocity of $\vmax \ge 130 \; \mathrm{km \; s^{-1}}$.  However, we also consider samples selected by their peak mass or $\vpeak$ value.  In those cases we choose our cuts so that 98.5$\%$  of the haloes have more than 100 particles and 99.5\% have more than 50 particles at $z=0$.  Thus any errors introduced by under-resolved haloes will contribute very little to the overall results. With those cuts we have $\mpeak \ge 5.0 \times10^{11} \; \hmsun$ and $\vpeak \ge 160 \; \mathrm{km \; s^{-1}}$. Table \ref{tab:cut} shows the values of these cuts and the resulting number of haloes when each cut is made.  One can see that for all cuts there are more than 25 million haloes in the 48 simulation boxes that we use.

\begin{figure} 
\centering
\includegraphics[width=0.5\textwidth]{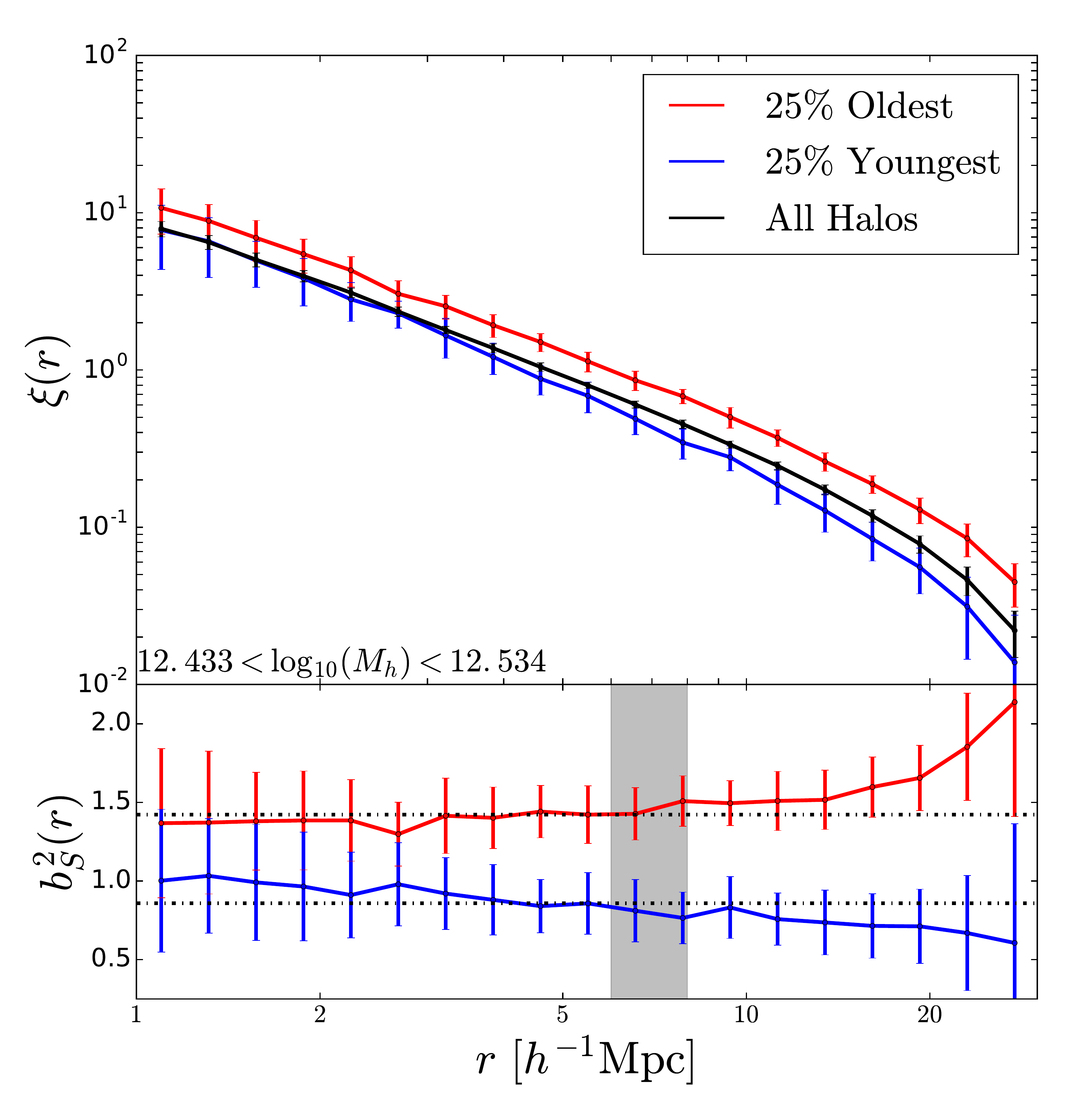}
\caption{Clustering of old vs. young haloes at fixed mass. The top panel shows the two-point correlation function of the oldest 25\% ({\it red line}), youngest 25\% ({\it blue line}), and all haloes ({\it black line}) in the narrow mass range $2.7-3.4\times 10^{12} \; \hmsun$. The bottom panel shows the square of the relative bias function for the 25\% oldest and youngest haloes, where the relative bias is defined by Equation~\ref{eqn:bias}. Dotted horizontal lines show the bias values for the two samples, averaged over all scales. In both panels, results are averaged over 48 simulation boxes and error bars show the uncertainly in the mean as estimated from the standard deviation among the boxes. Old haloes are clearly more clustered than young haloes at fixed mass. Moreover, we can see that the relative bias is weakly dependent on the distance between haloes.  We choose to focus on a range of 6-8$h^{-1}$ Mpc (illustrated by the shaded region), where the relative bias function is fairly flat.  We have checked both larger and smaller scales and find that none of our conclusions depend on the scales over which the relative bias is measured.}
\label{fig:cf}
\end{figure} 


\section{Relative Bias}
\label{sec:meth}

In order to study secondary bias we will be primarily looking at the relative bias of different subsets of haloes.  The relative bias is just the square root of the ratio of two correlation functions; we use {\sc{corrfunc}} \footnote{https://github.com/manodeep/Corrfunc} \citep[][]{corrfunc:17} to compute all correlation functions. One way of quantifying assembly bias is to measure the relative bias of a subset of haloes, selected by some halo property, compared to all haloes with the same mass.  For some property S this can be expressed as,

\beq
\label{eq:relbias}
b_{S}^{2}(r | \mhalo, S)={{\xi(r | \mhalo, S)}\over{\xi(r | \mhalo)}}.
\label{eqn:bias}
\eeq

Note that $b_{S}$ for all haloes of a given mass is $1.0$ by definition and is \emph{not} equal to the bias between those haloes and the dark matter distribution. Figure~\ref{fig:cf} shows the case of haloes in a narrow mass bin split into upper and lower quartiles by age.  The top panel shows the correlation functions computed for the oldest quartile, the youngest quartile, and all the haloes in the range $2.7-3.4 \times 10^{12} \;\hmsun$.  The bottom panel shows the square of the relative bias (as given by Eq.~\ref{eqn:bias}) calculated from these correlation functions.  Clearly, old haloes are more clustered than young haloes at fixed mass and this is precisely the phenomenon known as `halo assembly bias' \citep{st:04,gsw:05,harker:06,wech:06}. We see that the relative bias has very weak dependence on the distance between haloes.  Therefore, throughout this paper we consider clustering at only one length scale, $6-8 \; \h \; \mathrm{Mpc}$.  This scale is large enough to be safely in the 2-halo regime of $\xi(r)$ but is small enough that sample variance errors and finite box-size effects are negligible for this study. We have checked that our conclusions do not change if we instead adopt a smaller or larger separation. Although we do not study this further here, we note that the level of assembly bias does have a modest scale dependence, even above this scale.

\begin{figure*} 
\centering
\includegraphics[width=1.0\textwidth]{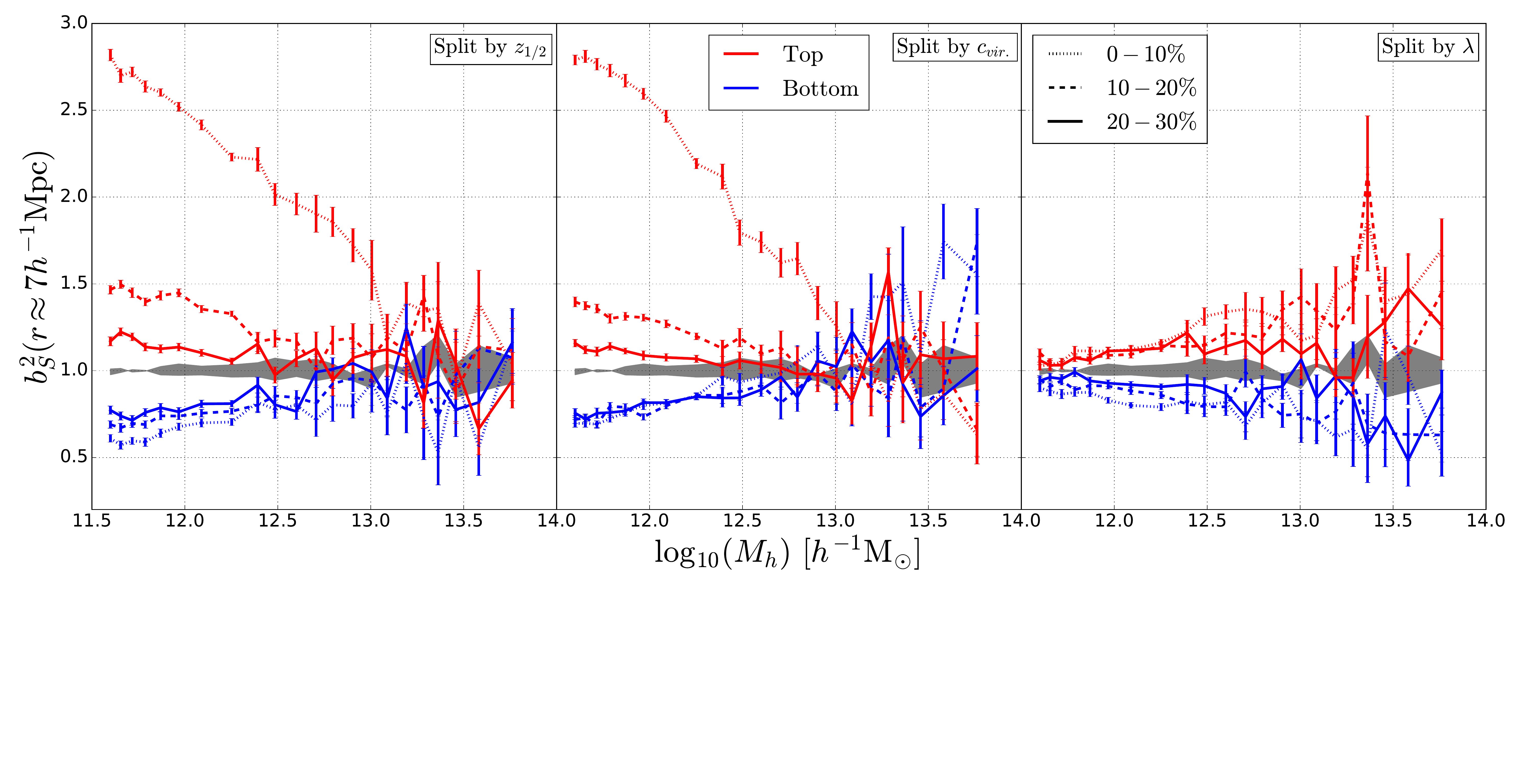}
\vspace{-2.5cm}
\caption{Relative bias of halo samples split by age ({\it left panel}), concentration ({\it middle panel}), and spin ({\it right panel}), as a function of halo mass. In each mass bin, we split the halo sample into deciles of each property and show the relative bias of the highest three deciles ({\it red lines}) and lowest three deciles ({\it blue lines}). For example, the dotted red curve in the left panel shows the correlation function of the 10\% oldest haloes divided by that of all haloes at a scale of $6-8\h$Mpc. Results are averaged over $\Nrun$ simulation boxes and error bars show the uncertainly in the mean as estimated from the standard deviation among the boxes. Gray shaded regions show the maximum effect associated with the finite width of the mass bins. The relative bias of the top decile of age or concentration is much higher than the other deciles, but this is not the case for spin nor for the bottom deciles. Halo assembly bias is thus largely driven by the high tail of the distribution when split by age or concentration. Assembly bias is a strong function of halo mass when split by age and concentration, but a weak function of halo mass that goes in the opposite direction when split by spin. As found by several previous studies, the sense of the bias switches at a halo mass of $\mhalo \approx 10^{13} \; \hmsun$ when split by concentration, but this does not occur for the halo mass range we probe when split by age.}
\label{fig:abias}
\end{figure*}

Figure~\ref{fig:abias} shows the square of the relative bias (measured at this scale) as a function of halo mass. In this case, we show results for the highest and lowest three deciles of halo age, concentration and spin. As can be seen clearly from the plot, when haloes of the same mass are split by these parameters there is a significant change in their clustering and this behavior is mass dependent. Age and concentration show many similar behaviors that are not shared when splitting the haloes by spin. For age and concentration, the bias of the top decile is much larger than that of the next decile for halo masses below $10^{13} \; \hmsun$; the three bottom deciles have similar bias.  For spin, the difference between the deciles is more symmetric and fairly small.  Based on the results for age and concentration, we see that most of the relative bias is coming from the top $20\%$ of haloes. We thus choose to focus on the top and bottom quartiles as this should conservatively account for most of the biased haloes.  We note that this isn't the case for spin, but we prefer to use one value for all properties, so we only consider quartiles for the rest of this paper.

As has been found by several past studies, the relative bias is a strong function of halo mass for age and concentration, with low mass haloes exhibiting the strongest assembly bias signal. In the case of concentration, the direction of the bias changes from low to high mass haloes, switching at around $10^{13} \; \hmsun$.  When split by age, however, there is no crossing in the halo mass range we probe, i.e., up to $5.0 \times 10^{13} \; \hmsun$. \citet{mzw:17} have shown that there is no age assembly bias for cluster mass haloes, although they find other forms of secondary bias. When split by spin, the halo mass dependence is much weaker and the sign is an increase of relative bias for more massive haloes \citep[this may be due to noisy measurements of halo spin at low masses; see for example,][]{benson:17}.  Thus the relative bias of haloes selected by their spin seems to be a different phenomenon than the case of age or concentration.  We show later that this continues to be the case when more halo properties are considered. Age and concentration are known to correlate with each other, with older haloes also having higher concentrations.  Thus if there is a bias arising from assembly history we would expect there to also be an induced assembly bias from concentration. However, the fact that the direction of the bias changes for concentration but not for age tells us that the concentration bias is not simply a consequence of age assembly bias combined with the correlation between age and concentration.  While there is some change in the age-concentration relation with halo mass, this change is modest, and the sign of the correlation remains the same. This result is also apparent for cluster mass haloes \citep{mzw:17}. 

The results in Figure~\ref{fig:abias} represent the mean of our $\Nrun$ simulation boxes, and the errors shown are the uncertainty in the mean, as estimated from the standard deviation among the boxes. Due to the large number of haloes in our simulations, these errors are quite small until we get to high mass systems. Another possible source of error is the finite width of our mass bins. Since age and concentration both correlate with halo mass, when we select older and more concentrated haloes we also preferentially select less massive haloes, which could be a problem since clustering depends on mass.  To quantify the possible contribution from this source of error we split each mass bin into the top and bottom quartiles of mass itself and calculate the relative bias of each. The range between these two biases is shown as the gray shaded region in Figure~\ref{fig:abias} and it thus represents the maximum possible deviation from $b=1$ that could be due to correlations of halo properties with halo mass. We will show this shaded region in all subsequent figures that display the relative bias, and we will refer to this uncertainty as the \emph{finite width error}. In Figure \ref{fig:abias} the measured relative bias is much larger than any contribution from this error for haloes with masses below $8 \times 10^{12} \; \hmsun$ for age and concentration and for all haloes when splitting by spin. 

\begin{figure*} 
\centering
\includegraphics[width=0.9\textwidth]{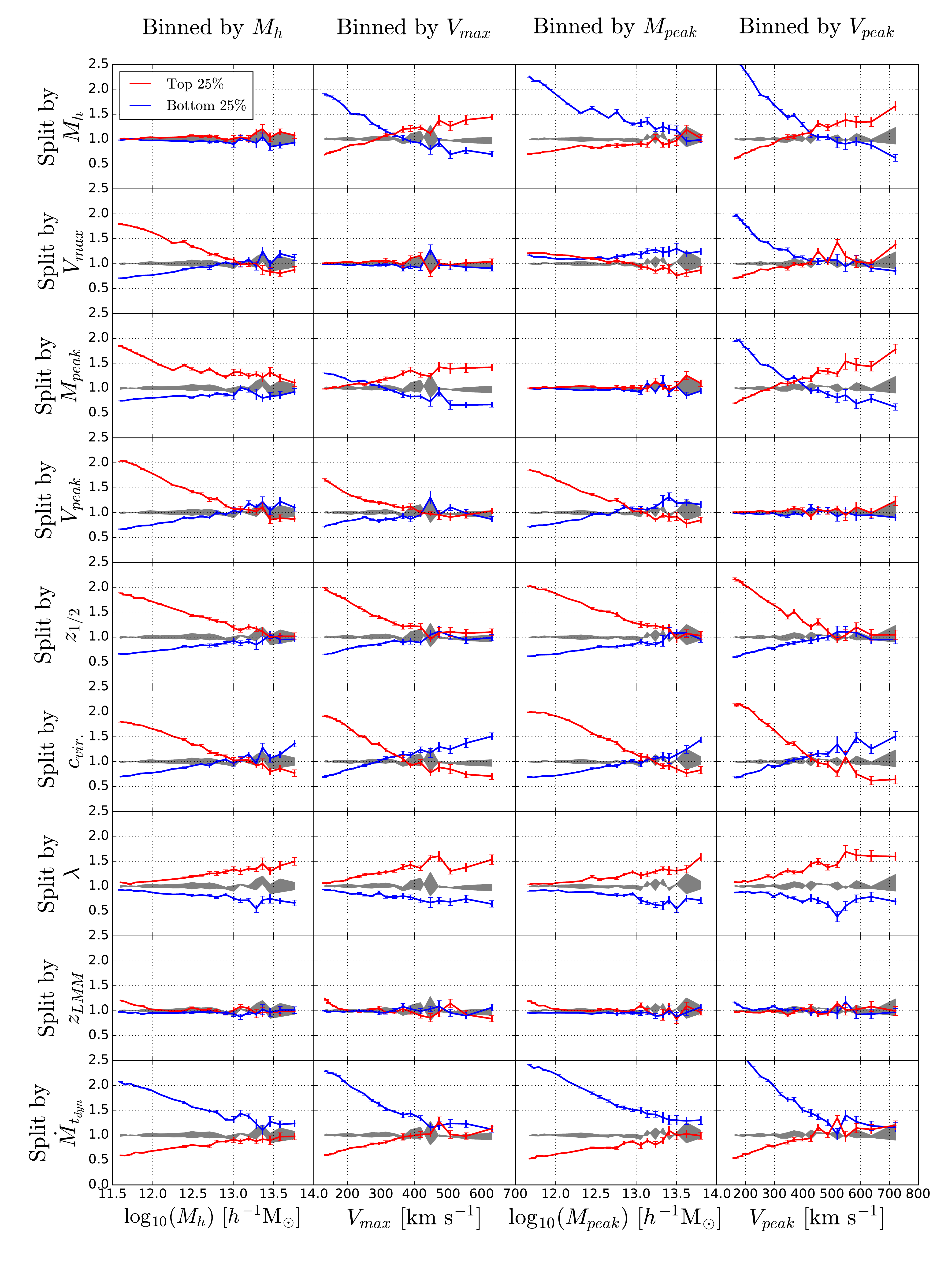}
\vspace{-1.0cm}
\caption{Relative bias of halo samples when binned by a primary mass-like property and split by a secondary property. The four columns of panels show the cases when haloes are binned by halo mass, maximum circular velocity, peak halo mass, and peak maximum circular velocity. Each row of panels then shows results when haloes are split by a different secondary property. In each panel, the red and blue lines show $b_{S}^2$ for the top and bottom quartiles of the secondary property, respectively. Results are always averaged over $\Nrun$ simulation boxes, and error bars show the uncertainly in the mean as estimated from the standard deviation among the boxes. Gray shaded regions show the finite width error, the maximum effect that would arise if the secondary parameter were perfectly correlated with the primary parameter. In the panels where the primary and secondary parameters are the same, this is the same as the measured bias.}
\label{fig:4x9}
\end{figure*}

\section{Secondary Bias}
\label{sec:split}

In this section we will consider a generalized formulation of assembly bias where we bin the haloes by one property and evaluate the relative bias when the haloes are split by a second property.  If the primary property the haloes are binned in is P and the secondary property they are split on is S then we can write the generalized relative bias as,
\beq
\label{eq:secbias}
b_{S}^{2}(r | P, S)={{\xi(r | P, S)}\over{\xi(r | P)}}.
\eeq
We see that equation \ref{eq:relbias} is just a special case of this where $P=\mhalo$.

When the primary property is something other than mass, we have to change how the sample of haloes is selected so that incompleteness does not affect the lowest bins.  Table~\ref{tab:cut} shows the selection made for each primary halo property to ensure that there is high completeness in all bins. In all the analysis that follows, halo mass or any other primary property is binned such that there are an equal number of haloes in each bin for low mass haloes.  For higher mass haloes this results in wide bins, making the \emph{finite width error} large, so we reduce the width of the bins until this error is smaller than the uncertainty calculated from the box-to-box variance.

It has become common to not only look at halo mass as the primary indicator of halo clustering, but other halo properties that strongly correlate with halo mass. Maximum circular velocity is often used instead of halo mass, and with a halo's history one can instead look at the peak value of halo mass or maximum circular velocity to try remove the effect of mass loss.  While we know haloes can be strongly stripped and often destroyed when they become subhaloes, it has also been shown that this mass loss can occur before the merger \citep{behr:14}.  In this section we explore whether the results of traditional assembly bias (as shown in Fig.~\ref{fig:abias}) where haloes are binned by mass hold up if haloes are instead binned by a different parameter.  We use four halo properties to bin the haloes: halo mass, peak halo mass, halo maximum circular velocity, and the peak maximum circular velocity. We refer to these as the primary property by which the haloes are binned and then we look at the relative bias if these bins are further split by a secondary property as described in equation \ref{eq:secbias}.  

Figure~\ref{fig:4x9} shows the matrix of results for these four primary properties (columns) split on nine secondary properties (rows). Each panel of the figure is similar to the panels shown in Figure~\ref{fig:abias}, except that here we only consider quartiles. We have looked at the deciles and the results are quite similar to what we found in Figure~\ref{fig:abias}, i.e., the assembly bias signal is about twice as high for the top decile but not much different for the bottom decile and there is little effect when the secondary property is halo spin. 

The first thing one notices is that there is a relative bias when haloes are split by a secondary property in all cases, except when the secondary property is $\zlmm$. In other words, a \emph{secondary bias} is not a particular feature of binning haloes by mass and splitting by age or concentration, but instead it is generic feature when looking at haloes binned by any quantity and then split by a second quantity\footnote{We have also examined the case of binning the haloes by age, concentration or spin and splitting by any other property (not shown here) and as might be expected still find a secondary bias exists.}. We see that irrespective of what primary property we bin on, the relative bias when split by age, concentration, spin, and mass accretion rate remains rather similar. The mass dependence is quite similar and the change of direction we saw for concentration also occurs if the primary parameter is $\vmax, \mpeak$ or $\vpeak$ (albeit at different mass scales). Now that we show quartiles rather than deciles, the signal to noise of these results is sufficiently high to see definitively that there is no corresponding change of 
direction in the case of halo age. This is especially evident when we bin by $\vmax$ or $\vpeak$ because in those cases the age assembly bias signal vanishes at a low enough velocity scale that we would easily see a change of direction if it existed. This reinforces the conclusion that secondary bias with concentration is distinct from age bias, at least to some extent.

Perhaps the most interesting aspect of Figure~\ref{fig:4x9} is the upper part of the figure where we use halo mass, maximum circular velocity, peak mass and peak maximum circular velocity as the secondary parameters.  We find that the relative bias when splitting on a mass-like secondary property is typically just as large as when splitting on age or concentration, which are more obviously connected to assembly history.  (The $\mpeak-\vmax$ combination is the exception, showing much smaller secondary bias.) However, it is important to recall that we compute secondary bias in bins of the primary quantity, and within such a bin the value of a second mass-like quantity may indeed depend strongly on assembly history.  For example, in a bin of $\vmax$ the less massive haloes must be more concentrated, so the relative bias at low $\vmax$ is $b_{S} > 1$ for the lowest $\mhalo$ quartile, even though halo bias increases with $\mhalo$ when $\vmax$ is not considered. Another novel result is the strong secondary bias with $\mpeak$ when binning by $\mhalo$: haloes that have lost mass are significantly more clustered than haloes that have not.

In some cases the correlation between halo properties leads to two panels with almost identical results \citep[although in general this is not necessarily the case;][]{mzw:17}.  For example, binning by halo mass and splitting by either $\vmax$ or $\cvir$ produce almost identical biases.  This makes sense since at fixed halo mass $\cvir$ and $\vmax$ are perfectly correlated, provided the haloes follow an NFW profile.  In most cases the connection of these relationships to halo assembly history is not very clear, one reason we refer to these relative biases as \emph{secondary bias}. Another interesting feature is that for some combinations, like binned by $\mpeak$ and split by $\vmax$, both the top and bottom quartile are more clustered than the mean for low mass haloes. The middle quartiles must be less clustered then, which suggests a non-monotonic relation between the secondary property and clustering. In several cases, when the bias of the top and bottom quartiles cross, it need not be at a value of one like it happens to be when the primary halo property is halo mass.

\begin{figure*} 
\centering
\includegraphics[width=0.7\textwidth]{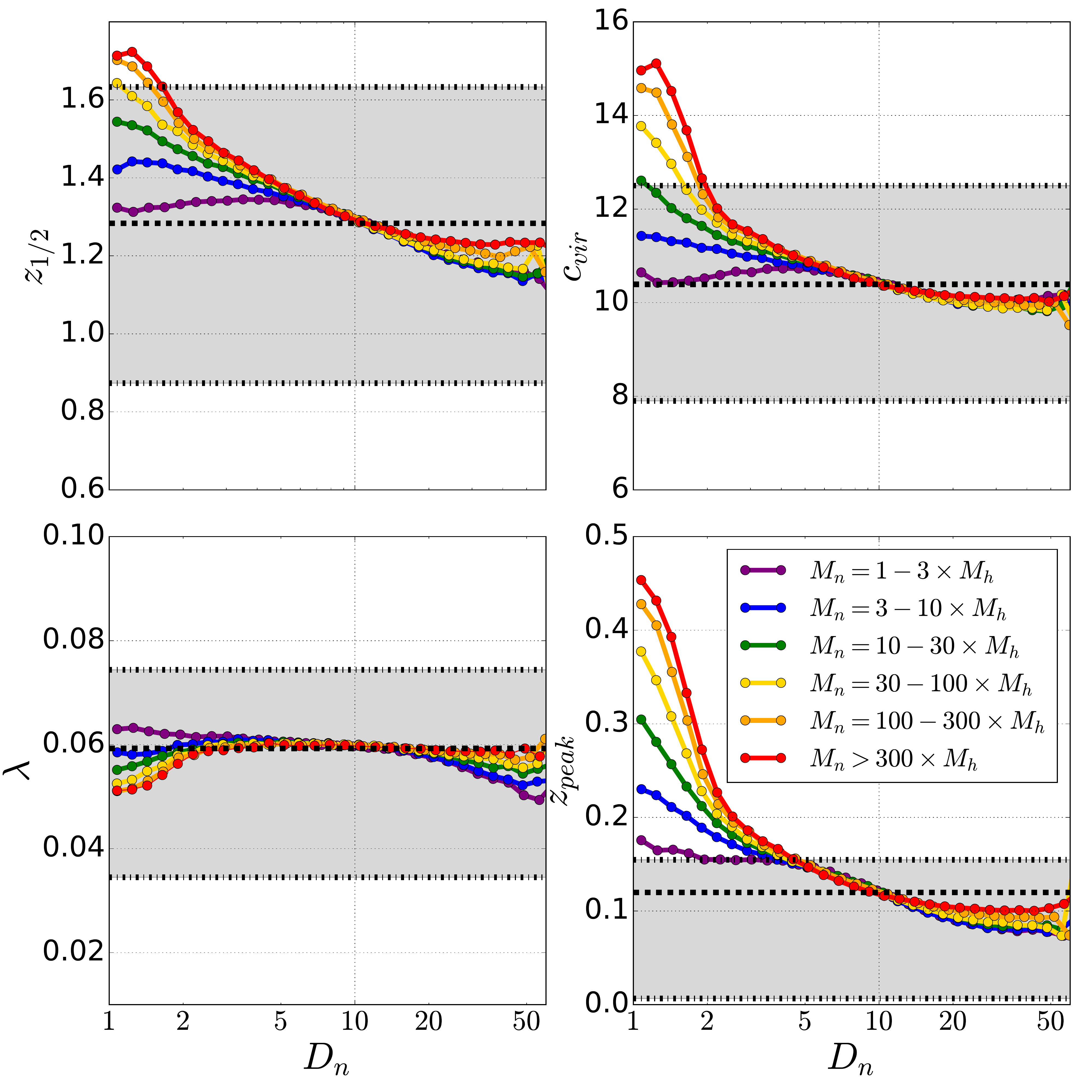}
\vspace{0.0cm}
\caption{The mean value of age ({\it top left panel}), concentration ({\it top right panel}), spin ({\it bottom left panel}) and redshift where the halo reached its peak mass ({\it bottom right panel}), as a function of closest distance to a more massive neighbour halo, $\Dn$, is shown for haloes in the mass range $3.76-4.07 \times 10^{11} \; \hmsun$. Different color lines show results for different definitions of what constitutes a more massive neighbour, as listed in the bottom right panel. The black dotted horizontal line in each panel shows the population average for that property, while the dot-dash lines bracket the middle 50\%. The distance $\Dn$ is normalized by the virial radius of the neighbour halo.}
\label{fig:neigh}
\end{figure*}

\section{Dependence of halo properties on neighbour distance}
\label{sec:Mc}

Early theoretical discussions of galaxy and halo bias focused on the biased clustering properties of the high peaks of a Gaussian field, the locations in the primordial fluctuations that would most naturally give rise to massive clusters and galaxies \citep{kaiser:84,bbks:86}.  The extended Press-Schechter or excurtion set formalism \citep{ps:74,bond:91,bower:91}, based on spherical collapse in the presence of a large scale background perturbation, provides a powerful tool for analytic modeling of halo bias \citep{mw:96}.  In the simplest version of this formalism, where pertubations from different scales are uncorrelated and the threshold overdensity for collapse $\delta_c$ is universal, the bias of haloes is predicted to depend strongly on halo mass but be independent of assembly history at fixed mass \citep{Wconf:96}. However, this prediction does not hold for more general assumptions \citep{zent:07}. Since the discovery of assembly bias in numerical simulations, numerous analytic and numerical studies have attempted to explain its origin. Some of these explanations emphasize properties of the initial fluctuations, departures from spherical collapse, and the influence of tidal fields \citep[e.g.][]{dala:08,desj:08,psd:13}. Others emphasize strongly non-linear effects \citep{wmj:07,wmj:09,dala:08,suna:16,borz:17,villarreal:17}, especially mass loss by haloes that have been stripped by passing through or near other haloes \citep[e.g][]{sinha:12}.  This stripping tends to increase a halo's inferred formation redshift, since it could reach 50\% of its $z=0$ mass relatively early, and it acts preferentially in the overdense environments around massive haloes.  Given the complex behavior of secondary biases shown in Figure~\ref{fig:4x9}, it is likely that more than one mechanism plays a significant role in assembly bias, and the importance of different mechanisms may be very different at masses $\mhalo \ll \mstar$ and $\mhalo > \mstar$.

\begin{figure*} 
\centering
\includegraphics[width=0.7\textwidth]{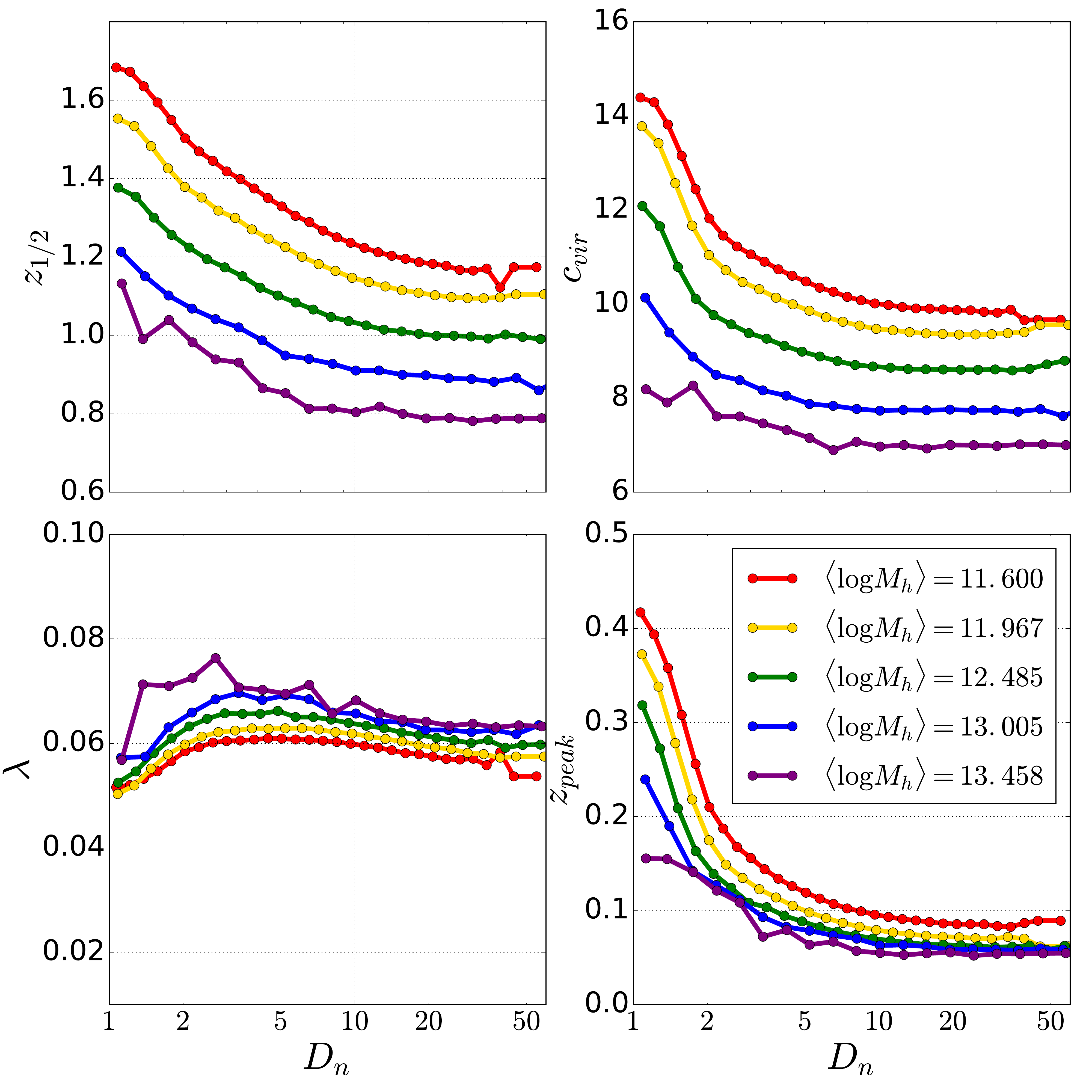}
\vspace{0.0cm}
\caption{The mean value of age ({\it top left panel}), concentration ({\it top right panel}), spin ({\it bottom left panel}) and redshift where the halo reached its peak mass ({\it bottom right panel}), as a function of closest distance to neighbour halo, $\Dn$, with at least ten times larger mass ($M_n > 10 \times \mhalo$). Different color lines show results for different halo mass bins, as listed in the top right panel. The distance $\Dn$ is normalized by the virial radius of the neighbour halo. A simple fitting function for the mean values of $\zhalf$ and $\cvir$ as a function of halo mass and $\Dn$ is given in the text.}
\label{fig:nmass}
\end{figure*}

In order to gain a better understanding of the origin of these secondary biases, we study how halo properties are influenced by their local environments at $z=0$.  This has been studied many times \citep{lk:99,bull:01b,st:04,avil:05,eshes:05, maul:07,macc:07,sm:11,lee:17}, usually by comparing halo properties to a measure of local density smoothed over some length scale. Instead, we compare halo properties to a halo's distance from a more massive neighbour.  This allows us to go to smaller scales because there is no need to smooth over a scale to define an environmental density. Also, while halo mass and environmental density are correlated, they are different things and it may be that the masses of individual haloes are more relevant here than the environmental density.  We define the normalized neighbour distance $\Dn$ of a given halo as 
\beq
\Dn = {D \over {R_{vir,n}}}
\eeq
where $D$ is the distance between the centres of the halo in question and the massive neighbour, and $R_{vir,n}$ is the neighbouring halo's virial radius. Using this normalized definition, the mass of the \emph{neighbouring} halo affects the distance such that a halo at the same physical distance is considered closer to a more massive neighbour than a less massive one. 

Figure~\ref{fig:neigh} shows how halo properties are correlated with their neighbour distance. Specifically, we take haloes in a narrow range of mass, $3.76-4.07 \times 10^{11} \; \hmsun$, and we show the mean value of their age, concentration, spin, and the redshift $\zpeak$ when they stopped accreting mass in bins of $\Dn$. We show results for different definitions of what constitutes a more massive neighbour, as listed in the legend. For example, to compute the red line in the top left panel, we perform the following steps: (1) consider each halo in the narrow mass range, (2) find its closest haloes of mass at least 300 times higher than itself, (3) calculate the distance to each neighbour in units of that halo's virial radius and set $\Dn$ equal to the smallest of these, (4) bin all the haloes in $\Dn$ and calculate the mean age in each bin. Figure~\ref{fig:neigh} reveals that the mean values of $\zhalf$, $\cvir$ and $\zpeak$ increase substantially for haloes that are close to a more massive neighbour. The amount of increase depends on how much more massive is the neighbour, with almost no effect for neighbours less than three times as massive as the halo, but a large effect for neighbours ten times as massive.  For spin there is very little dependence on distance to a massive neighbour, with only a slight decrease in the mean spin occurring for haloes within two virial radii of neighbours ten times or more as massive. The change in spin values are small compared to the overall spread of spin values (shown by the shaded region), but for age and concentration the change in the mean can exceed the halo-to-halo dispersion.     

\begin{figure*} 
\centering
\includegraphics[width=1.0\textwidth]{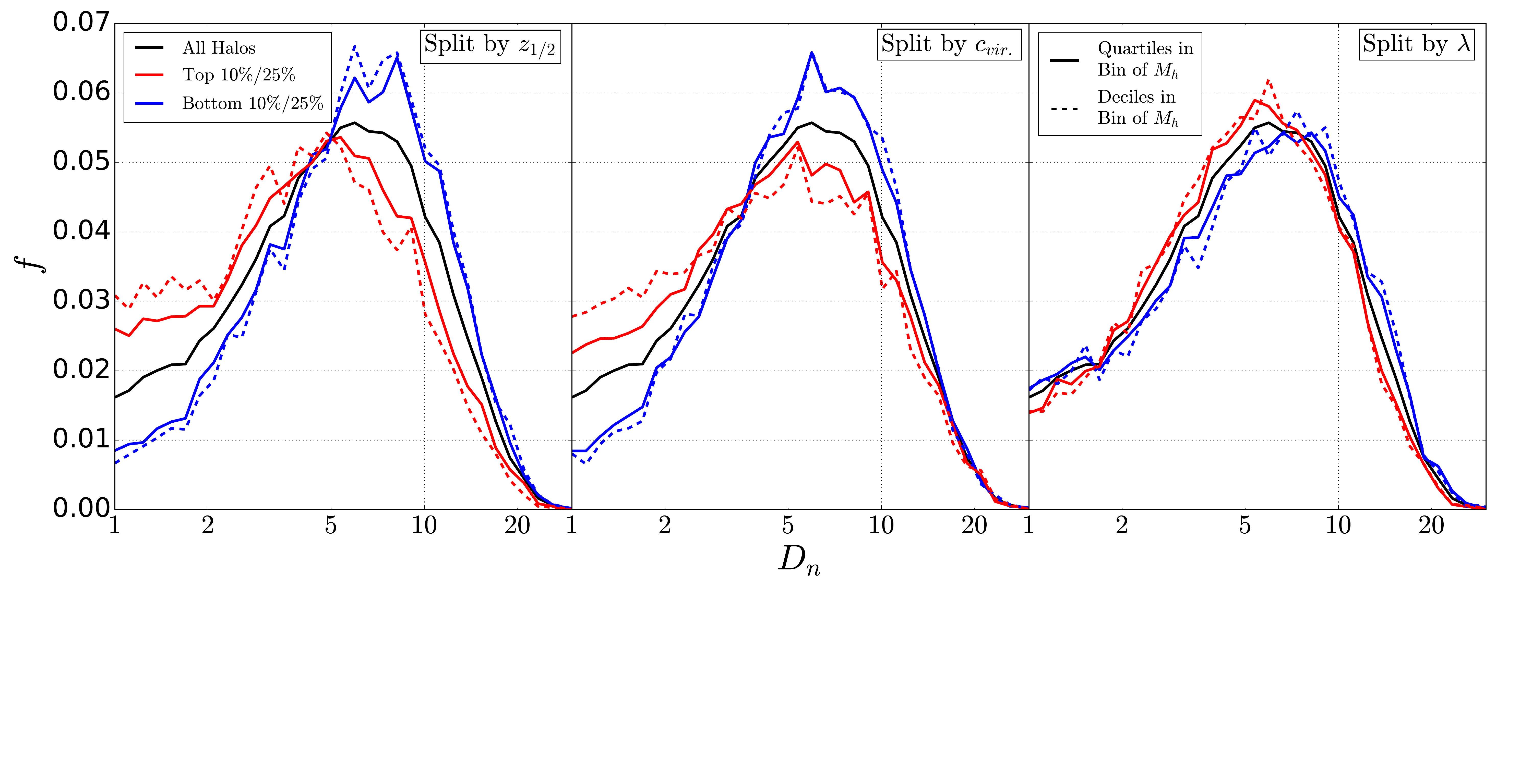}
\vspace{-2.8cm}
\caption{The distribution of halo distances $\Dn$ from a neighbouring halo that is at least ten times more massive ($M_n > 10 \times \mhalo$), for a samples of haloes in a narrow mass range ($3.76-4.07 \times 10^{11} \; \hmsun$) and further cut by age ({\it left panel}), concentration ({\it middle panel}), and spin ({\it right panel}). The black histogram corresponds to the full sample and is the same in all three panels, while red histograms show the top 10\% ({\it dashed}) and 25\% ({\it solid}) of each property and blue histograms show the bottom 25\% ({\it solid}) and 10\% ({\it dashed}) of each property.}
\label{fig:dist}
\end{figure*}

The mean values of $\zhalf$, $\cvir$ and $\zpeak$ exhibit a similar overall dependence on $\Dn$; they are highest right at the virial radius of the massive neighbour and they gradually drop with distance. The mean properties all converge to the global means (shown by the horizontal dashed lines) at $\Dn \sim 10$, and they drop below the mean at larger distances. For a halo that lives within a few virial radii of a massive neighbour, we might expect that the massive neighbour is directly influencing mass accretion onto that halo either by tidal forces or its effect on the velocity of dark matter particles around the halo. Additionally it has been shown that a significant fraction of these halos are ejected or `backsplash' halos that once resided within the virial radius of their massive neighbour \citep{wmj:09}. However, when $\Dn \gtrsim 5$ the massive neighbour cannot be directly influencing the halo, and the correlation is likely caused by a large scale environment that affects both the halo property and the probability of a massive neighbour. We see that for $\Dn$ greater than 10 the mean value of $\zhalf$, $\cvir$ and $\zpeak$ becomes less than the population average. At these distances haloes are in underdense regions which must be the cause of this small decrease in mean values for these properties.  Clearly the distance to a massive neighbour is highly correlated with other measures of environmental density; however, by not smoothing over a larger length scale we are able to see how significant this effect becomes for haloes that are very close to a massive neighbour and how it depends on the mass of that neighbour.  

\begin{figure*} 
\centering
\includegraphics[width=0.7\textwidth]{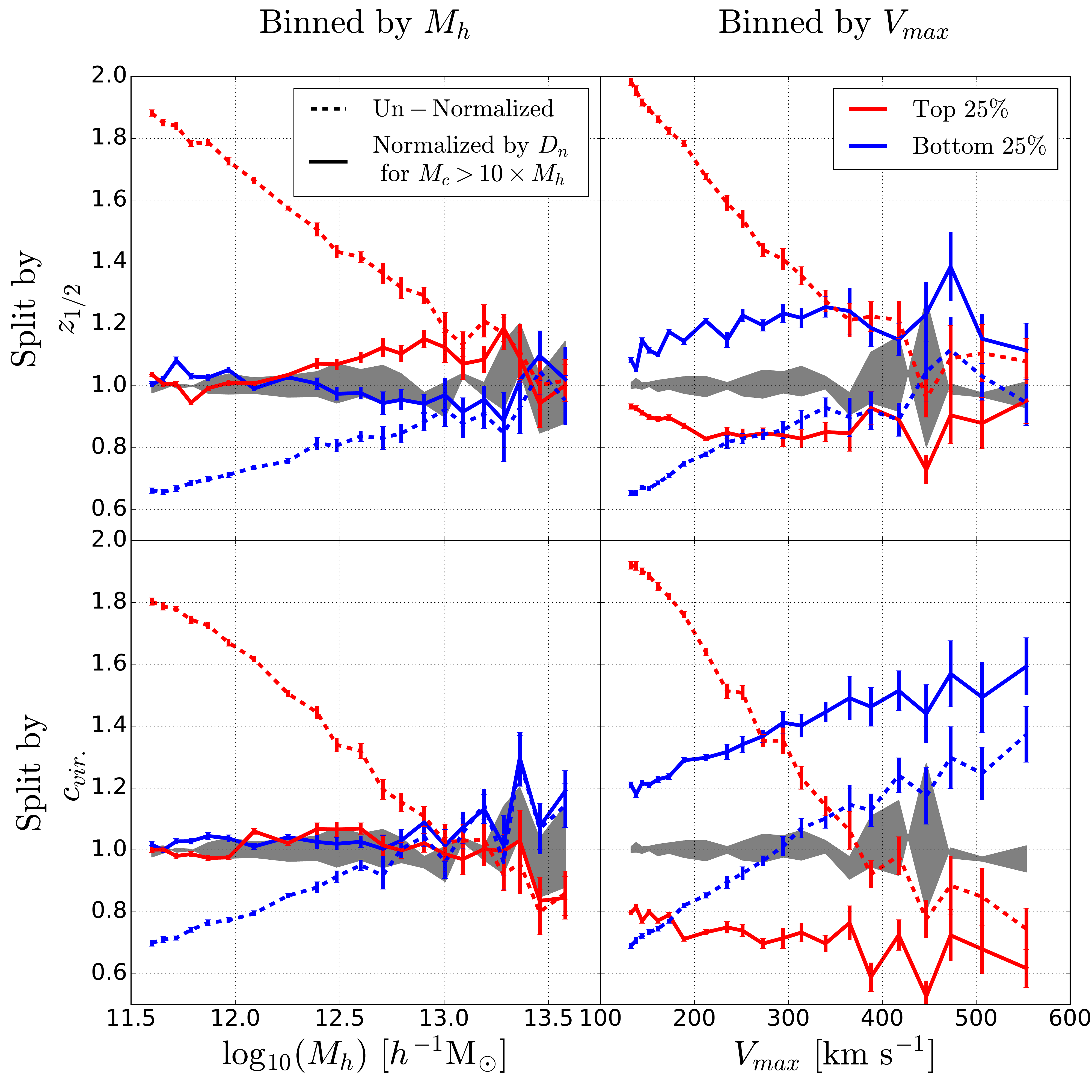}
\vspace{-0.2cm}
\caption{The relative bias when when the upper and lower quartiles of age ({\it top panels}) and concentration ({\it bottom panels}) are constructed to remove neighbour bias, i.e., to have identical distributions of distance from a more massive neighbour.  Left panels show binning by halo mass while right panels show  binning by maximum circular velocity. Dashed lines show the relative bias before removing neighbour bias, i.e., they are identical to the lines shown in the corresponding panels in Fig.~\ref{fig:4x9}, and solid lines show the result after controlling for neighbour bias. When binned by halo mass, removing neighbour bias largely eliminates the relative bias. In contrast, when binned by $\vmax$, there remains a large relative bias when split by age or concentration, but now the bias curves do not cross.}
\label{fig:norm}
\end{figure*}

The redshift where a halo reaches its peak mass, $\zpeak$, is the property most strongly affected by the distance to a massive neighbour.  The mean value of $\zpeak$ exceeds the one sigma deviation of the whole population for haloes with $\Dn < 5$, even when the neighbour is only 3 times more massive.  haloes that have stopped growing or have lost mass are predominantly near a more massive halo.  We can infer that this is the reason these haloes also have higher values of $\zhalf$ and $\cvir$.  Having reached their peak mass earlier, it is likely that these haloes have also reached half of their peak mass at an earlier time.  These haloes also tend to have density profiles that are steeper in the outer parts than NFW \citep[see for example,][]{lee:17}, which increases their concentrations (a steeper outer profile at fixed $\mhalo$ and $R_{vir}$ requires a smaller scale radius, $r_s$). Thus, proximity to a massive neighbour tends to lead to \emph{arrested development}, a higher value of $\zpeak$, which will cause such haloes to have higher values of age and concentration.  Many of these haloes have not only stopped growing, but have lost mass as well. For haloes with a $z=0$ halo mass of $\sim 10^{13} \; \hmsun$,  $15\%$ have lost at least $5\%$ of their peak mass value. For haloes with a mass of $\sim 10^{12} \; \hmsun$, $19\%$ have lost at least $5\%$ of their peak mass.  Mass loss for dark matter haloes is a much more common phenomenon \citep[][]{behr:14, lee:17} than commonly appreciated, a fact that is not accounted for in a spherical collapse type picture of halo formation.  

Figure~\ref{fig:neigh} shows that having a large change in the mean values of halo properties requires neighbours ten times as massive as the halo under consideration.  We thus adopt this neighbour mass ratio.  Using only these neighbours, $M_n > 10 \times \mhalo$, we now examine how the change in mean properties depends on halo mass. Figure~\ref{fig:nmass} shows clearly that the change is strongest for the lowest halo mass bin and decreases for higher mass haloes.  For the highest mass haloes there are not enough of them in the box to firmly say how their properties change, but we can see that the effect is trending towards no effect at higher masses.

The mean value of halo properties thus depends on the halo mass, the distance to a massive neighbour, and the ratio of that neighbour's mass to the halo mass.  We can fit a function to the mean age and concentration for neighbour mass ratios $M_n/\mhalo \ge 10$, as a function of halo mass and neighbour distance.  We find that the following function gives good fits to the curves in Figure~\ref{fig:nmass} over most values of $\Dn$:
\beq
\label{eq:fit}
\bar{p} = \Delta_p \left({{1 + D_p}\over{\Dn + D_p}}\right) + p_f ,
\eeq
where $p_f$ is the mean value of the property far from the massive neighbour, $\Delta_p$ is the maximum increase in the value of the property and $D_p$ is a scale in units of the neighbours virial radius which determines when the increase becomes significant.  Note that this function is a good fit over most of the range of $\Dn$ but not at $\Dn$ close to unity where the function tends to turn over. In that region it is better to evaluate equation \ref{eq:fit} at a $\Dn$ of 1.5 or 2 and use that value of $\bar{p}$ for smaller values of $\Dn$. For both $\cvir$ and $\zhalf$ we find that the parameters $\Delta_p, D_p$ and $p_f$ are well fit by power laws that depend only on the halo mass. For $\zhalf$ the best-fitting parameters are:

\begin{align*}
\Delta_{\zhalf} &= -0.186 \times \log \mhalo + 3.162, \\ 
D_{\zhalf} &= -0.163 \times \log \mhalo + 2.462, \\ 
{\zhalf}_f &= -0.186 \times \log \mhalo + 3.259
\end{align*}

and for $\cvir$ they are:

\begin{align*}
\Delta_{\cvir} &= -0.099 \times \log \mhalo + 0.837, \\
D_{\cvir} &= -2.147 \times \log \mhalo + 30.400, \\
{\cvir}_f &= -1.432 \times \log \mhalo + 26.222 
\end{align*}

These fitting functions should not be assumed to hold outside the range of halo masses we have probed, $\mhalo = 5 \times 10^{11} - 5 \times 10^{13} \; \hmsun$. 

The dependence of halo properties on distance from a massive neighbour provides a clear explanation of why some secondary bias occurs. If the mean value of a halo property depends on distance to a massive neighbour, then selecting haloes by that property will preferentially select haloes that are close to massive neighbours. These haloes will have \emph{neighbour bias}, that is, their clustering will reflect the mass of the massive neighbour they are close to instead of their own halo mass.  We can see how this happens in Figure~\ref{fig:dist}, which shows the distribution of $\Dn$ from a neighbour ten or more times as massive as the halo for all haloes in a mass bin, compared to that for a subset of the haloes selected by a secondary property.  We see that for age and concentration the distribution of $\Dn$ is strongly altered when haloes are selected by high or low percentiles of this property.  For spin, as we would expect based on Figures~\ref{fig:neigh} and~\ref{fig:nmass}, we see that the distribution of neighbour distances is only slightly changed when selecting the high or low values of spin.  In other words, halo spin shows little neighbour bias, unlike age and concentration, yet another example of how spin bias is different than other secondary biases.  For age and concentration, the distributions in Figure~\ref{fig:dist} are reminiscent of Figure~\ref{fig:abias}.  The bottom quartile and decile have essentially the same distribution, while the top decile has many more haloes with low $\Dn$ than the top quartile.  This is the same behavior of the relative bias in Figure~\ref{fig:abias}.  We have also looked at the distributions when binned by other properties (not shown) and found that the neighbour distributions change based on the primary property used to bin the haloes.  This sheds some light on why the relative biases in Figure~\ref{fig:4x9} can vary greatly when binned by a different primary property.  Clearly, neighbour bias contributes to some secondary biases and can explain some of the trends we have previously seen. 

To test whether neighbour bias is the sole driver of assembly bias, we can create a sample that has the same neighbour distance distribution, even when splitting it by age or concentration.  To do this we fit the top and bottom quartile as a function of neighbour distance with a fourth order polynomial. This is similar to equation~\ref{eq:fit} but a little more accurate.  We then determine which haloes are in the top and bottom quartiles of $\zhalf$ or $\cvir$ in bins of halo mass based on their values of $\Dn$. By construction this yields a sample that has no neighbour bias, but is split by age or concentration. Figure~\ref{fig:norm} shows the relative bias for such a sample. In the left panels we see that, when haloes are binned by mass, normalizing to a fixed $\Dn$ distribution drastically reduces the secondary bias, at least for haloes below $10^{13} \; \hmsun$. In other words, the presence of massive neighbours almost fully explains the secondary bias with $\zhalf$ or $\cvir$ at fixed $\mhalo$. Some effect remains at $\mhalo > 10^{13} \; \hmsun$, but our statistics are limited, and the error bars shown only come from the variance between boxes and do not include any error in our fit to the property dependence on $\Dn$. When binning by $\vmax$ (right panels), normalizing the $\Dn$ distribution does not remove the secondary bias with $\zhalf$ or $\cvir$, but it changes the behavior significantly. Most notably, after normalizing to the same $\Dn$ distribution it is the younger, less concentrated haloes that exhibit stronger clustering. Comparing the left and right panels of Figure~\ref{fig:norm} suggests that removing neighbour bias eliminates most assembly bias for $\mhalo$-selected halo samples but `overcorrects' for $\vmax$-selected samples. We do not show the analagous panels for bins of $\mpeak$ and $\vpeak$ but the effect of removing neighbour bias is similar to that seen in bins of $\mhalo$ and $\vmax$ respectively. 

\section{Summary and Discussion}
\label{sec:conc}

In this paper we have explored the clustering of dark matter haloes in a large suite of cosmological simulations with the goal of understanding how the clustering of haloes depends on their properties. In agreement with previous studies, we have found that assembly bias is a generic feature of dark matter halo clustering. More generally if haloes are selected to have one property fixed, then a subset of those selected by a second property usually exhibits biased clustering compared to the set of haloes in the first selection, which we refer to as secondary bias. The strength of the bias can be a strong function of the selectivity of the second property.  For example, haloes from the top $10\%$ of the concentration distribution are more than twice as biased as the next $10\%$ at fixed halo mass.  However, this behavior is not shared by all properties, and it is asymmetric; for the properties we have considered the relative bias of the \emph{bottom} three deciles is similar.

Exploring nine different halo properties, we find that binning haloes by one property and then selecting a subset by a second property generally gives a population with increased or decreased clustering.  This is the case even if the two properties are closely related, like halo mass and peak halo mass.  The strongest relative bias we find is for binning by peak maximum circular velocity and splitting by dynamical time averaged accretion rate.  The complexity of how halo clustering depends on different halo properties and how it varies with those properties suggests there is more than one underlying cause of these phenomena, a result also found by \citet{mzw:17} for cluster mass haloes.

A halo's proximity to a more massive neighbour can strongly influence the mean value of some halo properties like age, concentration and when a halo reaches its peak mass.  Proximity to a massive neighbour causes haloes to stop growing or even to lose mass. It may also cause haloes to grow slower than they would without the massive neighbour, though we have not examined this here. All of these effects can be grouped together as the \emph{arrested development} of the halo.  This arrested development causes the measured age and concentration of the haloes to be higher, such that when one selects the oldest or highest concentration haloes one is selecting haloes that are preferentially located close to much more massive haloes. As a result, the haloes inherit the higher clustering amplitude of their massive neighbours. If one controls for this neighbour bias, then the most common expression of assembly bias, binning by halo mass and splitting by age or concentration, is almost entirely removed.  We therefore conclude that these forms of clustering bias are caused by those halo properties being altered for haloes that are near massive neighbours. However, other forms of secondary bias, either splitting by a different property or binning by a different property, do not vanish when controlling for proximity to massive neighbours. Thus while neighbour bias explains some forms of secondary bias, removing it actually increases the secondary bias for other properties.

All of this leads to the conclusion that the correlation between halo properties and halo clustering is a complex phenomenon, with likely more than one physical cause. We summarize the types of bias discussed hear in as follows:

\begin{itemize}[leftmargin=*]
\item \emph{Secondary Bias} - When haloes are binned by any property there can exist a secondary bias when split by a different halo property.  Standard assembly bias where haloes are binned by mass and split by age, concentration, spin or another property is a particular example of this.  When the binning is done by another mass-like halo property we see that the resulting relative bias is very similar.  However, we find as strong or stronger effects when binning a sample by a one mass-like halo property and then splitting it by another mass-like property.  The one exception to this is peak halo mass and maximum circular velocity, which results in a fairly weak secondary bias. The only property we find that doesn't give a secondary bias is the redshift of the last major merger.
\item \emph{Neighbour Bias} - Many halo properties show a strong dependence on the distance to a more massive neighbour halo.  Thus selecting haloes by a high value of such a property can preferentially select haloes that are close to much more massive neighbours, resulting in neighbour bias.  This bias seems to completely explain the secondary bias of haloes binned by halo mass (or $\mpeak$) and split by age or concentration.  However, when haloes are binned by other mass like properties, like maximum circular velocity, and the neighbour bias is removed these haloes still show large secondary biases when split by age or concentration.  
\item \emph{Spin Bias} -  The secondary bias when haloes are split by spin behaves differently than when split by other halo properties.  There is almost no dependence on halo mass or other mass-like properties unlike the case for age and concentration. Moreover, the relative bias does not increase when looking at the top $10\%$ instead of top $25\%$.  Finally, spin has only a weak dependence on distance to massive neighbours, not enough to cause significant neighbour bias.  Together these findings suggest that the secondary bias from halo spin is caused by a separate physical mechanism than the other secondary biases we have explored. It is quantitatively less important, and the trends of spin with environment are weak compared to the intrinsic dispersion of the spin distribution, in agreement with older studies based on much smaller simulations \citep{be:87}.
\end{itemize}

The most common usage of the term `assembly bias' would be expressed in this language as secondary bias of haloes binned by halo mass and split by either age or concentration.  This bias can be explained largely by neighbour bias in the mass range considered here. However, we have also found a large number of secondary biases that are not explained by neighbour bias, including spin bias.  When the effect of neighbour bias is removed, the secondary bias of haloes binned by maximum circular velocity and split by age or concentration resembles that of spin bias. It is thus possible that spin bias is unusual in not having neighbour bias and that other secondary biases have a combination of neighbour bias and other causes of secondary bias.

One suggestion that has been made is that halo properties may be correlated with the density fluctuations in the linear regime that form the halo.  The height of the density peak or its curvature may partly determine a haloes age or concentration and thus correlations in the early density field can create secondary biases \citep{zent:07,desj:08,dala:08}.  It may be the case that early density fluctuations create an initial set of secondary biases and then later non-linear evolution (like arrested development) alters the initial mapping between halo properties and clustering.  This explains some aspects of secondary bias: why the relative bias for haloes binned by mass crosses when split by concentration but not when split by age.  Even when we account for neighbour bias, the concentration assembly bias signal remains for high-mass haloes.  Possibly its source is from early fluctuations that do not create a secondary bias when splitting by age. The strong dependence of some halo properties on neighbour distance within $\Dn = 1-3$ favors non-linear explanations of assembly bias, but the continuing dependence out to $\Dn = 10$ may be better explained by correlations rooted in initial conditions.

Regardless of physical origin, the fact that halo secondary bias is largely accounted for by neighbour bias has useful implications for modeling observations. First, it suggests focusing on distance to neighbouring groups and clusters when searching for signatures of {\it{galaxy}} assembly bias, as in some recent studies of `galaxy conformity' \citep{kauffmann:13}. Second, this description may be a good way to implement parametrized forms of assembly bias in HOD-based cosmological analyses, where the potential impact of assembly bias is treated via nuisance parameters.

The great complexity of assembly bias may actually be a good thing for simple models connecting galaxies to haloes.  If the relationship between halo properties and clustering depends on many properties in a complex way, then it is more likely that these effects may cancel out when translating to galaxy properties. This conjecture can be tested on semi-analytic galaxy formation models, which typically incorporate many effects and predict a multitude of galaxy observables. On the other hand, models that too tightly connect a galaxy property to only one halo property \citep[e.g. age-matching;][]{hw:13} may overestimate the secondary bias to be seen in galaxies \citep{para:15,lin:16,zm:16,mandelbaum:16}. Alternatively, one can test the validity of a single halo property strongly controlling a galaxy property by measuring the secondary bias in galaxy properties.  For example, galaxy disc size in most semi-analytic models is mostly determined by halo spin \citep[e.g.][]{some:12} and therefore should have a secondary bias like that found for spin.  If such a relationship is not found, then galaxy disc sizes are not primarily set by halo spin.  The study of secondary bias in the galaxy population can thus be a powerful tool to determine if and how halo properties influence galaxy properties.
  
\section*{Acknowledgements}
We thank Benjamin D. Wibking and Ying Zu for valuable comments on this work. The simulations used in this paper were produced by the LasDamas project (http://lss.phy.vanderbilt.edu/lasdamas/); we thank NSF XSEDE for providing the computational resources for LasDamas. Some of the computational facilities used in this project were provided by the Vanderbilt Advanced Computing Center for Research and Education (ACCRE). Figures in this paper were made with the matplotlib python package \citep{hunt:07}. A.N.S. was supported as an REU by grant NSF-REU 1263045 and is currently supported by the Department of Energy Computational Science Graduate Fellowship Program of the Office of Science and National Nuclear Security Administration in the Department of Energy under contract DE-FG02-97ER25308. A.A.B. and M.S. were supported by NSF Career Award AST-1151650. Parts of this research were conducted by the Australian Research Council Centre of Excellence for All Sky Astrophysics in 3 Dimensions (ASTRO 3D), through project number CE170100013. D.H.W. acknowledges support from NSF grant AST-1516997.

\bibliographystyle{mnras}
\bibliography{paper1}

\end{document}